\documentclass[12pt]{article}

\usepackage{amsmath}
\usepackage{amssymb}
\usepackage{amsfonts}
\usepackage [latin1]{inputenc}

\numberwithin{equation}{section}

\title{
Algebraic approach to a $d$-dimensional matrix Hamiltonian with so($d+1)$ symmetry}
\author{C. Quesne\thanks{e-mail: Christiane.Quesne@ulb.be}\\
{\small\sl D\'epartement de Physique,  Universit\'e Libre de Bruxelles,} \\ 
{\small\sl Campus de la Plaine CP229, Boulevard~du Triomphe, B-1050 Brussels, Belgium}}
\date{ }
\begin{document}
\baselineskip=22pt plus 1pt minus 1pt
\maketitle

\begin{abstract} 
A novel spin-extended so($d+1$,1) algebra is introduced and shown to provide an interesting framework for discussing the properties of a $d$-dimensional matrix Hamiltonian with spin 1/2 and so($d+1$) symmetry. With some $d+2$ additional operators, spanning a basis of an so($d+1$,1) irreducible representation, the so($d+1$,1) generators provide a very easy way for deriving the integrals of motion of the matrix Hamiltonian in Sturm representation. Such integrals of motion are then transformed into those of the matrix Hamiltonian in Schr\"odinger representation, including a Laplace-Runge-Lenz vector with spin. This leads to a derivation of the latter, as well as its properties in a more extended algebraic framework.
\end{abstract}


%
%
\newpage

\section{Introduction}

The hydrogen atom is known to have a nice symmetry (for a review see, e.g., \cite{wybourne}). Apart from the invariance of the Hamiltonian under rotations, leading to the conservation of the orbital angular momentum, it is also characterized by that of the Laplace-Runge-Lenz (LRL) vector, whose introduction in quantum mechanics dates back to Pauli \cite{pauli} and which was then studied by Fock \cite{fock}, Bargmann \cite{bargmann}, and many other authors. As a result, the hydrogen atom is a maximally superintegrable system, being characterized by an so(4), e(3), or so(3,1) invariance algebra for negative-, zero-, and positive-energy states, respectively. All these algebras may be embedded into an so(4,2) dynamical algebra \cite{barut71, barut79}.\par
%
%
More generally, the $d$-dimensional Coulomb problem, whose Hamiltonian is defined by $H = \frac{1}{2} {\bf p}^2 + \frac{\alpha}{r}$, where ${\bf p}^2 = \sum_{i=1}^d p_i^2$, $p_i = - {\rm i}\partial /\partial x_i$, $r = \left(\sum_{i=1}^d x_i^2\right)^{1/2}$, and $\alpha$ is some parameter, can be discussed in terms of an so($d+1$,2) Lie algebra and its bound states turn out to be a basis of an so($d+1$) algebra. As recently recalled \cite{cq}, this so($d+1$,2) algebra is a good starting point for discussing the Coulomb problem in Sturm representation, characterized by the operator $K = r \left(\frac{1}{2} {\bf p}^2 - E\right)$ and for deriving the invariance algebra of the latter in a very simple way. From these results, it is then straightforward to obtain the LRL vector corresponding to the Schr\"odinger representation, as well as its properties.\par
%
%
In the same study \cite{cq}, it was shown that such an approach can be applied to a generalization of the $d$-dimensional Coulomb problem to the $d$-dimensional Dunkl-Coulomb one, wherein the derivatives $\partial/\partial x_i$ are replaced by Dunkl operators $D_i$ \cite{dunkl89, dunkl14}. The latter are differential-difference operators, defined by $D_i = \partial/\partial x_i + (\mu_i/x_i) (1 - R_i)$, $i=1, 2, \ldots, d$, where $\mu_i$ is some positive parameter and $R_i$ is a reflection operator such that $R_i f(x_i) = f(-x_i)$. This has led to some deformed algebras and to a deformed LRL vector.\par
%
%
The purpose of the present paper is to apply the same method to another generalization of the $d$-dimensional Coulomb problem. Some years ago, a system describing a neutral particle with spin 1/2 and a non-trivial dipole momentum interacting with an external field inverse in radius in two \cite{pronko77} or three dimensions \cite{nikitin12} was shown to be endowed with some generalized LRL vector. A similar study was then carried out for arbitrary spin \cite{pronko07, nikitin13} and generalized to a $d$-dimensional space \cite{nikitin14}.\par
%
%
We plan to analyze along the lines of \cite{cq} the $d$-dimensional matrix Hamiltonian with spin 1/2 considered in \cite{nikitin14}. In Sect.~2, such a Hamiltonian is defined and some examples for low $d$ values are presented. In Sect.~3, a spin-extended so($d+1$,1) algebra is introduced together with some operators spanning an irreducible representation of the latter. In Sect.~4, such results are used to determine the invariance algebra of the Hamiltonian in Sturm representation. In Sect.~5, this invariance algebra is transformed into that in Schr\"odinger representation, thereby deriving the LRL vector with spin together with its properties. Finally, Sect.~6 contains the conclusion.\par
%
%
\section{\boldmath $d$-dimensional matrix problem with spin 1/2}

Let us consider the $d$-dimensional Hamiltonian \cite{nikitin14}
\begin{equation}
  H = \frac{1}{2} \mathbf{p}^2 + \frac{\alpha}{r^2} \boldsymbol{\gamma}\cdot \mathbf{x},  \label{eq:H} 
\end{equation}
where $\mathbf{p}^2 = \sum_{i=1}^d p_i^2$, $p_i = - {\rm i}\partial/\partial x_i$, $r = \left(\sum_{i=1}^d x_i^2\right)^{1/2}$, $\boldsymbol{\gamma} \cdot \mathbf{x} = \sum_{i=1}^d \gamma_i x_i$, and $\alpha$ is some constant. Here $\gamma_i$, $i=1, 2, \ldots, d$ are matrices that are basis elements of the Clifford algebra $\mathit{Cl}_d$, with defining relations
\begin{equation}
  \gamma_i \gamma_j + \gamma_j \gamma_i = 2 \delta_{i,j}.  \label{eq:gamma}
\end{equation}
In terms of these $\gamma_i$'s, the matrices
\begin{equation}
  S_{ij} = - \frac{\rm i}{4} (\gamma_i \gamma_j - \gamma_j \gamma_i). \label{eq:S}
\end{equation}
satisfy the so($d$) commutation relations
\begin{equation}
  [S_{ij}, S_{kl}] = {\rm i} \left(\delta_{i,k} S_{jl} + \delta_{i,l} S_{kj} + \delta_{j,k} S_{li} + \delta_{j,l} S_{ik} \right).
\end{equation}
As explained in \cite{nikitin14}, Hamiltonian (\ref{eq:H}) may be interpreted as describing a particle of spin 1/2.\par
%
%
We plan to deal with the Schr\"odinger equation
\begin{equation}
  H \Psi(\mathbf{x}) = E \Psi(\mathbf{x}),
\end{equation}
corresponding to (\ref{eq:H}), as well as the eigenvalue problem
\begin{equation}
  K \Psi(\mathbf{x}) = - \alpha \Psi(\mathbf{x}), \label{eq:Sturm}
\end{equation}
for the operator $K$ defined by
\begin{equation}
  K = (\boldsymbol{\gamma} \cdot \mathbf{x}) \left(\frac{1}{2}\mathbf{p}^2 - E\right) \label{eq:K}
\end{equation}
and corresponding to the so-called Sturm representation. Since from (\ref{eq:gamma}), it follows that
\begin{equation}
  (\boldsymbol{\gamma} \cdot \mathbf{x})^2 = r^2,  \label{eq:r^2}
\end{equation}
the operators $H$ and $K$ are related through the equations
\begin{equation}
  K = (\boldsymbol{\gamma} \cdot \mathbf{x})(H-E) - \alpha, \qquad H = \frac{\boldsymbol{\gamma} \cdot 
  \mathbf{x}}{r^2} (K+ \alpha) + E.  \label{eq:K-H}
\end{equation}
\par
%
%
It is worth observing that the total angular momentum operators $J_{ij} = L_{ij} + S_{ij}$, which generate an so($d$) algebra, commute with $H$ and $K$ because
\begin{equation}
  [J_{ij}, \boldsymbol{\gamma} \cdot \mathbf{x}] = \gamma_k [L_{ij}, x_k] + [S_{ij}, \gamma_k] x_k = 0
\end{equation}
since
\begin{equation}
  [L_{ij}, x_k] = - {\rm i} (\delta_{j,k} x_i - \delta_{i,k} x_j)
\end{equation}
and
\begin{equation}
  [S_{ij}, \gamma_k] = - {\rm i}(\delta_{j,k} \gamma_i - \delta_{i,k} \gamma_j).
\end{equation}
\par
%
%
{}For some low $d$ values, we may assume for instance the following $\gamma_i$'s:
\begin{equation}
\begin{split}
  & \bullet \quad d=2:  \gamma_1 = \sigma_1, \quad \gamma_2 = \sigma_2, \\
  & \bullet \quad d=3:  \gamma_1 = \sigma_1,  \quad \gamma_2 = \sigma_2, \quad \gamma_3 = \sigma_3, \\
  & \bullet \quad d=4:  \gamma_i = \begin{pmatrix}
        0 & {\rm i}\sigma_i \\
        -{\rm i}\sigma_i & 0
      \end{pmatrix}, \quad i = 1, 2, 3, \quad
      \gamma_4 = \begin{pmatrix}
        0 & I \\
        I & 0  
      \end{pmatrix},  \\
  & \bullet\quad d=5: \gamma_i = \begin{pmatrix}
        0 & {\rm i}\sigma_i \\
        -{\rm i}\sigma_i & 0
      \end{pmatrix}, \quad i = 1, 2, 3, \quad
      \gamma_4 = \begin{pmatrix}
        0 & I \\
        I & 0  
      \end{pmatrix},  \\
  &\hphantom{\bullet \quad d=5: {}} \; \; \gamma_5 = \begin{pmatrix}
       I & 0\\
       0 & -I
      \end{pmatrix}, 
\end{split}. \label{eq:gamma-examples}
\end{equation} 
where $\sigma_1$, $\sigma_2$, $\sigma_3$ denote the Pauli spin matrices and $I$ is the 2x2 unit matrix. For such choices, Hamiltonian (\ref{eq:H}) is a 2x2 or a 4x4 matrix Hamitonian if $d=2, 3$ or $d=4,5$, respectively.\par
%
%
The corresponding so($d$) generators (\ref{eq:S}) read
\begin{equation}
\begin{split}
  & \bullet \quad d=2:  S_{12} = \tfrac{1}{2} \sigma_3, \\
  & \bullet \quad d=3:  S_{ij} = \tfrac{1}{2} \epsilon_{ijk} \sigma_k, \quad i,j,k = 1,2,3, \\
  & \bullet \quad d=4:  S_{ij} = \tfrac{1}{2} \epsilon_{ijk} \begin{pmatrix}
        \sigma_k & 0 \\
        0 & \sigma_k
      \end{pmatrix}, \quad i, j, k = 1, 2, 3, \\
  & \hphantom{\bullet \quad d=5: {}} \; \:      S_{i4} = \tfrac{1}{2} \begin{pmatrix}
        \sigma_i & 0 \\
        0 & -\sigma_i  
      \end{pmatrix}, \quad i = 1, 2, 3,  \\
  & \bullet \quad d=5:  S_{ij} = \tfrac{1}{2} \epsilon_{ijk} \begin{pmatrix}
        \sigma_k & 0 \\
        0 & \sigma_k
      \end{pmatrix}, \quad i, j, k = 1, 2, 3, \\
  & \hphantom{\bullet \quad d=5: {}} \; \:      S_{i4} = \tfrac{1}{2} \begin{pmatrix}
        \sigma_i & 0 \\
        0 & -\sigma_i  
      \end{pmatrix}, \quad i = 1, 2, 3,  \\
  & \hphantom{\bullet \quad d=5: {}} \; \:      S_{i5} = - \tfrac{1}{2} \begin{pmatrix}
        0 & \sigma_i \\
        \sigma_i & 0  
      \end{pmatrix}, \quad i = 1, 2, 3,  \\
  & \hphantom{\bullet \quad d=5: {}} \; \:      S_{45} = \tfrac{{\rm i}}{2} \begin{pmatrix}
        0 & I \\
        - I & 0  
      \end{pmatrix},  \\
\end{split}. \label{eq:S-examples}
\end{equation}
where $\epsilon_{ijk}$ stands for the totally antisymmetric tensor with summation over repeated indices.\par
%
%
\section{\boldmath Spin-extended so($d+1$,1) algebra}

\setcounter{equation}{0}

Let us now introduce $\tfrac{1}{2}(d+1)(d+2)$ operators defined by
\begin{equation}
  \begin{split}
  J_{ij} &= L_{ij} + S_{ij} = - J_{ji}, \qquad L_{ij} = x_i p_j - x_j p_i, \\
  A_i &= \frac{1}{2} x_i \mathbf{p}^2 - \left(\mathbf{x} \cdot \mathbf{p} - {\rm i}\frac{d-1}{2}\right) p_i
     - \frac{1}{2}x_i + S_{ij} p_j, \\ 
  M_i &= \frac{1}{2} x_i \mathbf{p}^2 - \left(\mathbf{x} \cdot \mathbf{p} - {\rm i}\frac{d-1}{2}\right) p_i
     + \frac{1}{2}x_i + S_{ij} p_j, \\
  T &= \mathbf{x} \cdot \mathbf{p} - {\rm i} \frac{d-1}{2}, 
  \end{split} \label{eq:so}
\end{equation}
where $i$ and $j$ run over 1, 2, \ldots, $d$, $\mathbf{x} \cdot \mathbf{p} = \sum_{i=1}^d x_i p_i$, and there is a summation over repeated indices.\par
%
%
It is straightforward to show that these operators satisfy the following commutation relations
\begin{equation}
\begin{split}
  &[J_{ij}, J_{kl} ] = {\rm i}\left(\delta_{i,k} J_{jl} + \delta_{i,l} J_{kj} + \delta_{j,k} J_{li} + \delta_{j,l} J_{ik}\right), 
       \\
  &[J_{ij}, A_k] = {\rm i} \left(\delta_{i,k} A_j - \delta_{j,k} A_i \right), \\
  &[J_{ij}, M_k] = {\rm i} \left(\delta_{i,k} M_j - \delta_{j,k} M_i \right), \\
  &[J_{ij}, T] = 0, \\
  &[A_i, A_j] = - [M_i, M_j] = {\rm i} J_{ij}, \\
  &[A_i, M_j ] = {\rm i} \delta_{i,j} T, \\
  &[A_i, T] = - {\rm i} M_i, \qquad [M_i, T] = - {\rm i} A_i.  
\end{split}. \label{eq:so-com}
\end{equation}
\par
%
%
Here, as in the case where $S_{ij} = 0$, these spin-dependent operators are the generators ${\cal L}_{ab} = - {\cal L}_{ba} = {\cal L}_{ab}^{\dagger}$ of an so($d+1, 1$) algebra, whose commutation relations are given by
\begin{equation}
  [{\cal L}_{ab}, {\cal L}_{cd}] = {\rm i} (g_{ac} {\cal L}_{bd} + g_{ad} {\cal L}_{cb} + g_{bc} {\cal L}_{da} +
  g_{bd} {\cal L}_{ac}),
\end{equation}
where $g_{ab} = {\rm diag}(1, 1, \ldots, 1, -1)$. The identifications are the following ones:
\begin{equation}
\begin{split}
  & {\cal L}_{ij} = J_{ij}, \qquad i, j = 1, 2, \ldots, d, \\
  & {\cal L}_{i, d+1} = A_i, \qquad {\cal L}_{i,d+2} = M_i, \qquad i=1,2, \ldots, d, \\
  & {\cal L}_{d+1,d+2} = T.
\end{split}
\end{equation}
\par
%
%
The second-order Casimir operator of this algebra reads
\begin{equation}
  Q_2 = \mathbf{J}^2 + \mathbf{A}^2 - \mathbf{M}^2 - T^2.  \label{eq:Q_2-def}
\end{equation}
As shown in appendix A, it reduces to the constant
\begin{equation}
  Q_2 = - \tfrac{1}{8} (d-1) (d+2),  \label{eq:Q_2}
\end{equation}
instead of $Q_2 = - \frac{1}{4} (d-1) (d+1)$, obtained in the $S_{ij} = 0$ case.\par
%
%
Let us now introduce $d+2$ additional operators, defined by
\begin{equation}
\begin{split}
  &\Gamma_0 = \tfrac{1}{2} (\boldsymbol{\gamma} \cdot \mathbf{x})(\mathbf{p}^2 + 1), \\
  &\Gamma_{d+1} = \tfrac{1}{2} (\boldsymbol{\gamma} \cdot \mathbf{x})(\mathbf{p}^2 - 1), \\
  &\Gamma_i = (\boldsymbol{\gamma} \cdot \mathbf{x}) p_i, \qquad i = 1, 2, \ldots, d. \label{eq:Gamma}
\end{split}
\end{equation}
As in the standard $d$-dimensional Coulomb problem, where $\boldsymbol{\gamma} \cdot \mathbf{x}$ is replaced by $r$ in (\ref{eq:Gamma}) and $S_{ij} = 0$ in (\ref{eq:so}) (see \cite{wybourne}), they form an irreducible representation of so($d+1, 1$).Their commutation relations with operators (\ref{eq:so}) are indeed given by
\begin{equation}
\begin{split}
  &[J_{ij}, \Gamma_k] = {\rm i}(\delta_{i,k} \Gamma_j - \delta_{j,k} \Gamma_i), \\
  &[A_i, \Gamma_{d+1}] = - [M_i, \Gamma_0] = - {\rm i} \Gamma_i, \\
  &[T, \Gamma_0] = {\rm i} \Gamma_{d+1}, \qquad [T, \Gamma_{d+1}] = {\rm i} \Gamma_0, \\
  &[J_{ij}, \Gamma_0] = [J_{ij}, \Gamma_{d+1}] = [A_i, \Gamma_0] = [M_i, \Gamma_{d+1}] = [T, \Gamma_i]
      =0.
\end{split}. \label{eq:so-Gamma}
\end{equation}
 \par
 %
 %
In contrast with the Coulomb case, however, the commutators of operators $\Gamma_0$, $\Gamma_{d+1}$, and $\Gamma_i$ with one another do not give back operators $J_{ij}$, $A_i$, $M_i$, and $T$, so that the set of operators (\ref{eq:so}) and (\ref{eq:Gamma}) does not close an so($d+1, 2$) algebra. One indeed gets the following results:
\begin{equation}
\begin{split}
  &[\Gamma_0, \Gamma_{d+1}] = {\rm i} \left(T + {\rm i} \frac{d-1}{2} + {\rm i} L_{ij} S_{ij}\right), \\
  &[\Gamma_i, \Gamma_0] = - {\rm i} M_i - S_{ij} x_j(\mathbf{p}^2 + 1) - \left(\frac{d-1}{2} + L_{jk} S_{jk}
      \right)p_i + {\rm i} S_{ij} p_j, \\
  &[\Gamma_i, \Gamma_{d+1}] = - {\rm i} A_i - S_{ij} x_j(\mathbf{p}^2 + 1) - \left(\frac{d-1}{2} + L_{jk} S_{jk}
      \right)p_i + {\rm i} S_{ij} p_j, \\ 
  &[\Gamma_i, \Gamma_j] = - {\rm i} J_{ij} + {\rm i} S_{ij} - 2 x_k(S_{ik}p_j - S_{jk}p_i),
\end{split}
\end{equation}
where use has been made of the relations
\begin{equation}
  (\boldsymbol{\gamma} \cdot \mathbf{x}) \gamma_i = x_i - 2{\rm i}S_{ij}x_j, \qquad (\boldsymbol{\gamma}
  \cdot \mathbf{x}) (\boldsymbol{\gamma} \cdot \mathbf{p}) = \mathbf{x} \cdot \mathbf{p} + {\rm i}
  L_{ij} S_{ij},  \label{eq:relations}
\end{equation}
resulting from (\ref{eq:gamma}) and (\ref{eq:S}).\par
%
%
It is also worth observing that $\Gamma_0$, $\Gamma_{d+1}$, and $T$, which, in the Coulomb case, close an so(2,1) algebra with a Casimir operator given by \cite{cq}
\begin{equation}
  \Gamma_0^2 - \Gamma_{d+1}^2 - T^2 = \mathbf{J}^2 + \tfrac{1}{4}(d-1)(d-3),  \label{eq:casimir}
\end{equation}
do not satisfy such a property here. Nevertheless, as shown in appendix A, a relation rather similar to (\ref{eq:casimir}) is obtained in the present case, namely
\begin{equation}
  \Gamma_0^2 - \Gamma_{d+1}^2 - T^2 = \mathbf{J}^2 + \frac{1}{8}(d-1)(d-2).  \label{eq:Casimir}
\end{equation}
\par
%
%
\section{Invariance algebra of the $d$-dimensional matrix problem in Sturm representation}

\setcounter{equation}{0}

{}From its definition (\ref{eq:K}), it is obvious that the operator $K$, which is diagonalized in Sturm representation, is a linear combination of operators $\Gamma_0$ and $\Gamma_{d+1}$, defined in (\ref{eq:Gamma}),
\begin{equation}
  K = \tfrac{1}{2}(1-2E) \Gamma_0 + \tfrac{1}{2}(1+2E) \Gamma_{d+1}.
\end{equation}
It therefore directly follows from (\ref{eq:so-Gamma}) that the operators $J_{ij}$ and $B_i$, defined by
\begin{equation}
  B_i = \tfrac{1}{2}[(1-2E) A_i + (1+2E) M_i],
\end{equation}
are such that
\begin{equation}
  [J_{ij}, K] = [B_i, K] = 0.
\end{equation}
Hence, these $\frac{1}{2}d(d+1)$ operators are integrals of motion of the $d$-dimensional matrix problem in Sturm representation, described by equation (\ref{eq:Sturm}).\par
%
%
{}From the definitions of $A_i$ and $M_i$, given in (\ref{eq:so}), the explicit form of $B_i$ is
\begin{equation}
  B_i = \frac{1}{2} x_i \mathbf{p}^2 - \left(\mathbf{x} \cdot \mathbf{p} - {\rm i}\frac{d-1}{2}\right) p_i
  + S_{ij} p_j + E x_i.  \label{eq:B}
\end{equation}
It is also a direct consequence of (\ref{eq:so-com}) that the integrals of motion $J_{ij}$ and $B_i$ satisfy the commutation relations
\begin{equation}
\begin{split}
  &[J_{ij}, J_{kl}] = {\rm i} (\delta_{i,k} J_{jl} + \delta_{i,l} J_{kj} + \delta_{j,k} J_{li} + \delta_{j,l} J_{ik}), \\
  &[J_{ij}, B_k] = {\rm i} (\delta_{i,k} B_j - \delta_{j,k} B_i), \\
  &[B_i, B_j] = -2 {\rm i} E J_{ij}.
\end{split}. \label{eq:com-J-B}
\end{equation}
\par
%
%
In the Coulomb case, the operators $J_{ij}$ and $B_i$ satisfy some additional relations, given in (4.7) and (4.8) of \cite{cq} for $\mu_i=0$. Let us first consider the counterpart of (4.8), connecting $\mathbf{B}^2$, $K^2$, and $\mathbf{J}^2$.\par
%
%
{}From (\ref{eq:B}), we may split $B_i$ into a spin-independent term and a spin-dependent one, as follows:
\begin{equation}
  B_i = B_i^{(1)} + B_i^{(2)},
\end{equation}
with
\begin{equation}
\begin{split}
  &B_i^{(1)} = \frac{1}{2} x_i \mathbf{p}^2 - \left(\mathbf{x} \cdot \mathbf{p} - {\rm i}\frac{d-1}{2}\right) p_i
       + E x_i, \\
  &B_i^{(2)} = S_{ij} p_j.
\end{split}
\end{equation}
Hence
\begin{equation}
  \mathbf{B}^2 = \left(\mathbf{B}^{(1)}\right)^2 + \mathbf{B}^{(1)} \cdot \mathbf{B}^{(2)} + \mathbf{B}^{(2)}
  \cdot \mathbf{B}^{(1)} + \left(\mathbf{B}^{(2)}\right)^2,
\end{equation}
where
\begin{align}
  \left(\mathbf{B}^{(1)}\right)^2 &= \frac{1}{4} \Bigl\{r^2 \mathbf{p}^4 - 2{\rm i}\Bigl(\mathbf{x} \cdot
       \mathbf{p} - {\rm i} \frac{d-1}{2}\Bigr) \mathbf{p}^2 + 4E \Bigl[r^2 \mathbf{p}^2 - 2(\mathbf{x} \cdot
       \mathbf{p})^2 \nonumber \\
   & \quad {} + {\rm i} (2d-3) \mathbf{x} \cdot \mathbf{p} + \frac{1}{2}d(d-1)\Bigr] + 4E^2 r^2\Bigr\},
\end{align}
as a special case of (4.10) of \cite{cq}. In addition, it is straightforward to show that
\begin{equation}
  \mathbf{B}^{(1)} \cdot \mathbf{B}^{(2)} + \mathbf{B}^{(2)} \cdot \mathbf{B}^{(1)} = \frac{1}{2} L_{ij}
  S_{ij} \left(\mathbf{p}^2 + 2E\right)
\end{equation}
and
\begin{equation}
  \left(\mathbf{B}^{(2)}\right)^2 = S_{ij} S_{ik} p_j p_k = \frac{1}{2} \{S_{ij}, S_{ik}\} p_j p_k
  = \frac{1}{4} (d-1) \mathbf{p}^2,
\end{equation}
where use has been made of
\begin{equation}
  \{S_{ij}, S_{ik}\} = \frac{1}{2} (d-1) \delta_{j,k},
\end{equation}
directly deriving from (\ref{eq:gamma}) and (\ref{eq:S}). The explicit expression of $\mathbf{B}^2$ is therefore
\begin{align}
  \mathbf{B}^2 &= \frac{1}{4} \Bigl\{r^2 \mathbf{p}^4 - 2{\rm i} (\mathbf{x} \cdot \mathbf{p}) \mathbf{p}^2
       + 4E\Bigl[r^2 \mathbf{p}^2 - 2(\mathbf{x} \cdot \mathbf{p})^2 + {\rm i}(2d-3) \mathbf{x} \cdot \mathbf{p}
       \nonumber \\
  & \quad {}+\frac{1}{2}d(d-1) \Bigr] + 4E^2 r^2\Bigr\} + \frac{1}{2} L_{ij} S_{ij} (\mathbf{p}^2 + 2E).
       \label{eq:B^2}
\end{align}
\par
%
%
On the other hand, $K^2$ is easily calculated by using the relation $[\mathbf{p}^2, \boldsymbol{\gamma} \cdot \mathbf{x}] = - 2{\rm i} \boldsymbol{\gamma} \cdot \mathbf{p}$, as well as equation (\ref{eq:relations}). The result reads
\begin{align}
  K^2 &= \frac{1}{4} r^2 \mathbf{p}^4 - \frac{{\rm i}}{2} (\mathbf{x} \cdot \mathbf{p}) \mathbf{p}^2 +
      \frac{1}{2} L_{ij} S_{ij} \mathbf{p}^2 - E (r^2 \mathbf{p}^2 - {\rm i} \mathbf{x} \cdot \mathbf{p}
      + L_{ij} S_{ij}) \nonumber \\
  & \quad {}+ E^2 r^2.
\end{align}
\par
%
%
With $\mathbf{J}^2$ given in equation (\ref{eq:term-1}), it is then obvious that equation (4.8) of \cite{cq} is replaced by
\begin{equation}
  \mathbf{B}^2 = K^2 + 2E \left[\mathbf{J}^2 + \frac{1}{8}d(d-1)\right].
\end{equation}
\par
%
%
Next, let us consider the counterpart of equation (4.7) of \cite{cq}. It is not difficult to see that here $J_{ij} B_k + J_{jk} B_i + J_{ki} B_j$ does not vanish for $1 \le i < j < k \le d$ as in the previous case. In the three-dimensional case, however, one can get an interesting result for $\mathbf{J} \cdot \mathbf{B}$, where $J_i = \frac{1}{2} \epsilon_{ijk} J_{jk}$ and similarly $L_i = \frac{1}{2} \epsilon_{ijk} L_{jk}$, $S_i = \frac{1}{2} \epsilon_{ijk} S_{jk}$, satisfying the commutation relations
\begin{equation}
  [J_i, J_j] = {\rm i} \epsilon_{ijk} J_k, \qquad [L_i, L_j] = {\rm i} \epsilon_{ijk}  L_k, \qquad [S_i, S_j] =
  {\rm i} \epsilon_{ijk} S_k.
\end{equation}
From
\begin{equation}
\begin{split}
  &\mathbf{L} \cdot \mathbf{B}^{(1)} = 0, \\
  &\mathbf{L} \cdot \mathbf{B}^{(2)} = (\mathbf{x} \cdot \mathbf{p}) (\mathbf{p} \cdot \mathbf{S}) -
       (\mathbf{x} \cdot \mathbf{S}) \mathbf{p}^2, \\
  &\mathbf{S} \cdot \mathbf{B}^{(1)} = \tfrac{1}{2} (\mathbf{x} \cdot \mathbf{S}) \mathbf{p}^2
       -(\mathbf{x} \cdot \mathbf{p} - {\rm i}) (\mathbf{p} \cdot \mathbf{S}) + E \mathbf{x} \cdot \mathbf{S}, \\
  &\mathbf{S} \cdot \mathbf{B}^{(2)} = - {\rm i} \mathbf{p} \cdot \mathbf{S},
\end{split}
\end{equation}
one indeed obtains
\begin{equation}
  \mathbf{J} \cdot \mathbf{B} = - (\mathbf{x} \cdot \mathbf{S}) (\tfrac{1}{2} \mathbf{p}^2 - E).
  \label{eq:J-B}
\end{equation}
In the special case where one assumes $\gamma_i = \sigma_i$ as in (\ref{eq:gamma-examples}), and therefore $\mathbf{S} = \frac{1}{2}\boldsymbol{\sigma} = \frac{1}{2} \boldsymbol{\gamma}$ according to (\ref{eq:S-examples}), equation (\ref{eq:J-B}) becomes
\begin{equation}
  \mathbf{J} \cdot \mathbf{B} = - \tfrac{1}{2} K.  \label{eq:J-B-bis}
\end{equation}
\par
%
%
\section{\boldmath Invariance algebra of the $d$-dimensional matrix problem in Schr\"odinger representation}

\setcounter{equation}{0}

On considering $H$ instead of $K$, it is obvious that the components of the angular momentum operator $\mathbf{J}$ remain integrals of the motion:
\begin{equation}
  [J_{ij}, H] = 0.
\end{equation}
This is not the case, however, for the components of $\mathbf{B}$, which have therefore to be transformed. We plan to show that the operators $\tilde{A}_i$, defined by
\begin{equation}
  \tilde{A}_i = B_i + x_i (H-E),  \label{eq:LRL}
\end{equation}
have such a property, namely
\begin{equation}
  [\tilde{A}_i, H] = 0,  \label{eq:LRL-H}
\end{equation}
and are therefore a generalization of the LRL vector components to the present matrix problem.\par
%
%
{}From (\ref{eq:LRL}) and (\ref{eq:LRL-H}), this amounts to showing that
\begin{equation}
  [B_i, H] + [x_i, H] (H-E) = 0.  \label{eq:LRL-proof}
\end{equation}
Since
\begin{equation}
  [x_i, H] = \left[x_i, \frac{1}{2} \mathbf{p}^2\right] = {\rm i} p_i,  \label{eq:commutator}
\end{equation}
and
\begin{equation}
  [B_i, H] = \left[B_i, \frac{\boldsymbol{\gamma} \cdot \mathbf{x}}{r^2}\right] (K+\alpha)
  = \left[B_i, \frac{\boldsymbol{\gamma} \cdot \mathbf{x}}{r^2}\right] (\boldsymbol{\gamma} \cdot \mathbf{x})
  (H-E),
\end{equation}
where use has been made of (\ref{eq:K-H}), equation (\ref{eq:LRL-proof}) can be transformed into
\begin{equation}
  \left[B_i, \frac{\boldsymbol{\gamma} \cdot \mathbf{x}}{r^2}\right] (\boldsymbol{\gamma} \cdot \mathbf{x})
  = - {\rm i} p_i 
\end{equation}
or, equivalently,
\begin{equation}
  \left[B_i, \frac{\boldsymbol{\gamma} \cdot \mathbf{x}}{r^2}\right] = - {\rm i} p_i 
  \frac{\boldsymbol{\gamma} \cdot \mathbf{x}}{r^2}.  \label{eq:LRL-proof-bis}
\end{equation}
The proof of this equation is detailed in appendix B.\par
%
%
We conclude that the components of the spin-extended LRL vector can be written as
\begin{equation}
  \tilde{A}_i = x_i \mathbf{p}^2 - \left(\mathbf{x} \cdot \mathbf{p} - {\rm i} \frac{d-1}{2}\right) p_i + S_{ij} p_j
  + \alpha x_i \frac{\boldsymbol{\gamma} \cdot \mathbf{x}}{r^2}
\end{equation}
and coincide with the operators $K_{\mu}$ defined in equation (5) of \cite{nikitin14}.\par
%
%
The commutation relations of the integrals of motion $J_{ij}$ and $\tilde{A}_i$ in Schr\"odinger representation among themselves can be directly derived from those of the integrals of motion $J_{ij}$ and $B_i$ in Sturm representation, given in (\ref{eq:com-J-B}), and are given by
\begin{equation}
\begin{split}
  &[J_{ij}, J_{kl}] = {\rm i} (\delta_{i,k} J_{jl} + \delta_{i,l} J_{kj} + \delta_{j,k} J_{li} + \delta_{j,l} J_{ik}), \\
  &[J_{ij}, \tilde{A}_k] = {\rm i} (\delta_{i,k}\tilde{A}_j - \delta_{j,k} \tilde{A}_i), \\
  &[\tilde{A}_i, \tilde{A}_j] = - 2{\rm i} H J_{ij}.
\end{split}
\end{equation}
\par
%
%
To prove the last relation, use is made of
\begin{equation}
  [\tilde{A}_i, \tilde{A}_j] = [B_i + x_i (H-E), B_j + x_j (H-E)],
\end{equation}
where $[B_i, B_j]$ is already known from (\ref{eq:com-J-B}),
\begin{equation}
  [B_i, x_j (H-E)] - [B_j, x_i (H-E)] = (-2{\rm i} J_{ij} + {\rm i} L_{ij})(H-E),  \label{eq:B-xH}
\end{equation}
and
\begin{equation}
  [x_i(H-E), x_j(H-E)] = - {\rm i} L_{ij} (H-E).
\end{equation}
The demonstration of (\ref{eq:B-xH}) is based upon the relations
\begin{equation}
  [B_i, x_j (H-E)]= \left[B_i, x_j \frac{\boldsymbol{\gamma} \cdot \mathbf{x}}{r^2}\right] (K+\alpha)
  = \left[B_i, x_j \frac{\boldsymbol{\gamma} \cdot \mathbf{x}}{r^2}\right] (\boldsymbol{\gamma} \cdot \mathbf{x})
  (H-E), 
\end{equation}
\begin{equation}
  [B_i, x_j] = [\tfrac{1}{2}(1-2E) A_i + \tfrac{1}{2}(1+2E) M_i, M_j - A_j] = {\rm i} \delta_{i,j} T - {\rm i} J_{ij},
  \label{eq:B-x}
 \end{equation}
and equation (\ref{eq:LRL-proof-bis}).\par
%
%
In appendix C, $\mathbf{\tilde{A}}^2$ is shown to be expressible in terms of $H$ and $\mathbf{J}^2$ as
\begin{equation}
  \mathbf{\tilde{A}}^2 = 2H \left(\mathbf{J}^2 + \frac{1}{8}d(d-1)\right) + \alpha^2.  \label{eq:relation-1} 
\end{equation}
This equation slightly differs from the corresponding equation (5.19) for the Coulomb problem, obtained in \cite{cq}.\par
%
%
A counterpart of equation (\ref{eq:J-B-bis}), obtained for $d=3$ and $\gamma_i = \sigma_i$, can also be easily derived and is given by
\begin{equation}
  \mathbf{J} \cdot \mathbf{\tilde{A}} = \tfrac{1}{2} \alpha.  \label{eq:relation-2}
\end{equation}
Here use is made of (\ref{eq:LRL}), (\ref{eq:J-B-bis}), and $\mathbf{J} \cdot \mathbf{x} = \frac{1}{2} 
\boldsymbol{\sigma} \cdot \mathbf{x}$.\par
%
%
It is worth observing that equations (\ref{eq:relation-1}) and (\ref{eq:relation-2}) coincide with some results obtained in \cite{nikitin13} for $d=3$ by direct calculations. Equation (\ref{eq:relation-1}), valid for any $d$, is however a novel result.\par
%
%
\section{Conclusion}

In the present work, we have shown that a known $d$-dimensional matrix Hamiltonian with spin 1/2 can be analyzed in the framework of a novel spin-extended so($d+1$,1) algebra.\par
%
%
We have introduced $d+2$ additional operators, which span a basis of an so($d+1$,1) irreducible representation. Although they do not close an so($d+1$,2) algebra with the so($d+1$,1) generators, as their counterparts for the Coulomb problem, they are essential to easily derive the integrals of motion of the matrix Hamiltonian in Sturm representation.\par
%
%
Such an invariance algebra can then be transformed into that of the matrix Hamiltonian in Schr\"odinger representation. This provides us with a derivation of the LRL vector with spin, as well as the properties of the latter, in a more extended algebraic framework.\par
%
%
Analyzing the $d$-dimensional matrix Hamiltonian with higher spin in the same kind of algebraic framework would be an interesting subject for future work.\par
%
%
\section*{Acknowledgements}

This work was supported by the Fonds de la Recherche Scientifique - FNRS under Grant Number 4.45.10.08.\par
%
%
\section*{Data availability statement}

No new data were created or analyzed in this study.\par
%
%
\section*{Appendix A. Proof of equations (\ref{eq:Q_2}) and (\ref{eq:Casimir})}

\renewcommand{\theequation}{A.\arabic{equation}}
\setcounter{equation}{0}

Let us start with the proof of equation (\ref{eq:Q_2}). From (\ref{eq:so}), we successively obtain
\begin{equation}
  \mathbf{J}^2 = \tfrac{1}{2} J_{ij} J_{ij} = \tfrac{1}{2}(L_{ij} L_{ij} + 2 L_{ij} S_{ij} + S_{ij} S_{ij}),
\end{equation}
where
\begin{equation}
  \tfrac{1}{2} L_{ij} L_{ij} = r^2 \mathbf{p}^2 - (\mathbf{x} \cdot \mathbf{p})^2 + {\rm i} (d-2) \mathbf{x}
  \cdot \mathbf{p}
\end{equation}
and
\begin{equation}
  \tfrac{1}{2} S_{ij} S_{ij} = \tfrac{1}{8} d(d-1),
\end{equation}
resulting from (\ref{eq:gamma}) and (\ref{eq:S}). Hence,
\begin{equation}
  \mathbf{J}^2 = r^2 \mathbf{p}^2 - (\mathbf{x} \cdot \mathbf{p})^2 + {\rm i} (d-2) \mathbf{x} \cdot
  \mathbf{p} + L_{ij} S_{ij} + \tfrac{1}{8} d(d-1).  \label{eq:term-1}
\end{equation}
\par
%
%
On other hand,
\begin{align}
  \mathbf{A}^2 - \mathbf{M}^2 &= - \left\{\frac{1}{2} x_i \mathbf{p}^2 - \left(\mathbf{x} \cdot \mathbf{p}
         - {\rm i} \frac{d-1}{2}\right) p_i + S_{ij} p_j\right\} x_i \nonumber \\
  & \quad {} - x_i \left\{\frac{1}{2} x_i \mathbf{p}^2 - \left(\mathbf{x} \cdot \mathbf{p}
         - {\rm i} \frac{d-1}{2}\right) p_i + S_{ij} p_j\right\},
\end{align}
from which we get
\begin{equation}
  \mathbf{A}^2 - \mathbf{M}^2 = - r^2 \mathbf{p}^2 + 2 (\mathbf{x} \cdot \mathbf{p})^2 - {\rm i} (2d-3)
  \mathbf{x} \cdot \mathbf{p} - L_{ij} S_{ij} - \tfrac{1}{2} d(d-1).  \label{eq:term-2}
\end{equation}
Furthermore
\begin{equation}
  T^2 = (\mathbf{x} \cdot \mathbf{p})^2 - {\rm i} (d-1) \mathbf{x} \cdot \mathbf{p} - \tfrac{1}{4} (d-1)^2.
  \label{eq:term-3}
\end{equation}
Inserting (\ref{eq:term-1}), (\ref{eq:term-2}), and (\ref{eq:term-3}) in (\ref{eq:Q_2-def}) leads to (\ref{eq:Q_2}), which is therefore proved.\par
%
%
On considering next equation (\ref{eq:Casimir}), we note that
\begin{align}
  \Gamma_0^2 - \Gamma_{d+1}^2 &= \tfrac{1}{4} (\boldsymbol{\gamma} \cdot \mathbf{x}) (\mathbf{p}^2
       +1) (\boldsymbol{\gamma} \cdot \mathbf{x}) (\mathbf{p}^2 + 1) - \tfrac{1}{4} (\boldsymbol{\gamma} \cdot 
       \mathbf{x}) (\mathbf{p}^2 -1) (\boldsymbol{\gamma} \cdot \mathbf{x}) (\mathbf{p}^2 - 1) \nonumber \\
   &= (\boldsymbol{\gamma} \cdot \mathbf{x})^2 \mathbf{p}^2 - {\rm i} (\boldsymbol{\gamma} \cdot \mathbf{x})
        (\boldsymbol{\gamma} \cdot \mathbf{p}) \nonumber \\
   &= r^2 \mathbf{p}^2 - {\rm i} \mathbf{x} \cdot \mathbf{p} + L_{ij} S_{ij},
\end{align}
where use has been made of (\ref{eq:r^2}) and (\ref{eq:relations}). On combining this relation with (\ref{eq:term-3}), we get
\begin{equation}
  \Gamma_0^2 - \Gamma_{d+1}^2 - T^2 = r^2 \mathbf{p}^2 - (\mathbf{x} \cdot \mathbf{p})^2 + {\rm i} (d-2)
  \mathbf{x} \cdot \mathbf{p} + L_{ij} S_{ij} + \tfrac{1}{4} (d-1)^2.
\end{equation}
Comparison with (\ref{eq:term-1}) leads to equation (\ref{eq:Casimir}), which is therefore proved.
\par
%
%
\section*{Appendix B. Proof of equation (\ref{eq:LRL-proof-bis})}

\renewcommand{\theequation}{B.\arabic{equation}}
\setcounter{equation}{0}

To prove equation (\ref{eq:LRL-proof-bis}), we successively get the following results;
\begin{equation}
  \left[p_i, \frac{\boldsymbol{\gamma} \cdot \mathbf{x}}{r^2}\right] = - \frac{{\rm i}}{r^2} \gamma_i + 2{\rm i}
  \frac{x_i}{r^4} (\boldsymbol{\gamma} \cdot \mathbf{x}),
\end{equation}
\begin{equation}
  \left[\mathbf{p}^2, \frac{\boldsymbol{\gamma} \cdot \mathbf{x}}{r^2}\right] = - \frac{2{\rm i}}{r^2}
  (\boldsymbol{\gamma} \cdot \mathbf{p}) + \frac{4{\rm i}}{r^4} (\boldsymbol{\gamma} \cdot \mathbf{x})
  \left(\mathbf{x} \cdot \mathbf{p} - {\rm i} \frac{d-2}{2}\right),  \label{eq:B.2}
\end{equation}
\begin{equation}
  \left[\mathbf{x} \cdot \mathbf{p}, \frac{\boldsymbol{\gamma} \cdot \mathbf{x}}{r^2} \right] = 
  \frac{{\rm i}}{r^2} (\boldsymbol{\gamma} \cdot \mathbf{x}),  \label{eq:B.3}
\end{equation}
\begin{align}
  \left[\left(\mathbf{x} \cdot \mathbf{p} - {\rm i} \frac{d-1}{2}\right) p_i, \frac{\boldsymbol{\gamma} \cdot   
        \mathbf{x}}
       {r^2}\right] &= \left[- \frac{{\rm i}}{r^2} \gamma_i + 2{\rm i} \frac{x_i}{r^4} (\boldsymbol{\gamma} \cdot
       \mathbf{x})\right] \left(\mathbf{x} \cdot \mathbf{p} - {\rm i} \frac{d-5}{2}\right) \nonumber \\
   & \quad {}+ \frac{{\rm i}}{r^2} (\boldsymbol{\gamma} \cdot \mathbf{x}) p_i,
\end{align}
\begin{align}
  \left[S_{ij} p_j, \frac{\boldsymbol{\gamma} \cdot \mathbf{x}}{r^2}\right] &= S_{ij} \left[- \frac{{\rm i}}{r^2}
       \gamma_j + 2{\rm i} \frac{x_j}{r^4} (\boldsymbol{\gamma} \cdot \mathbf{x})\right] + \frac{{\rm i}}{r^2}
       [x_i (\boldsymbol{\gamma} \cdot \mathbf{p}) - \gamma_i (\mathbf{x} \cdot \mathbf{p})] \nonumber \\
   & = - \frac{{\rm i}}{r^2} \gamma_i \left(\mathbf{x} \cdot \mathbf{p} - {\rm i} \frac{d-3}{2}\right)
        - \frac{x_i}{r^4} (\boldsymbol{\gamma} \cdot \mathbf{x}) + {\rm i} \frac{x_i}{r^2} (\boldsymbol{\gamma}
        \cdot \mathbf{p}).
\end{align}
In the last equation, use is made of
\begin{equation}
  S_{ij} \gamma_j = - \frac{{\rm i}}{2} (d-1) \gamma_i, \qquad 
  S_{ij} x_j = - \frac{{\rm i}}{2}  [\gamma_i (\boldsymbol{\gamma} \cdot \mathbf{x}) - x_i], \label{eq:B.6}
\end{equation}
resulting from (\ref{eq:gamma}) and (\ref{eq:S}).\par
%
%
{}From (\ref{eq:B}), it therefore results that
\begin{equation}
  \left[B_i, \frac{\boldsymbol{\gamma} \cdot \mathbf{x}}{r^2}\right] = 2 \frac{x_i}{r^4} (\boldsymbol{\gamma}
  \cdot \mathbf{x}) - \frac{1}{r^2} \gamma_i - \frac{{\rm i}}{r^2} (\boldsymbol{\gamma} \cdot \mathbf{x})p_i,
\end{equation}
which coincides with the right-hand side of (\ref{eq:LRL-proof-bis}), thus completing the proof of the latter.\par
%
%
\section*{Appendix C. Proof of equation (\ref{eq:relation-1})}

\renewcommand{\theequation}{C.\arabic{equation}}
\setcounter{equation}{0}

To prove equation (\ref{eq:relation-1}), let us start from (\ref{eq:LRL}) and write $\mathbf{\tilde{A}}^2$ as
\begin{equation}
  \mathbf{\tilde{A}}^2 = \mathbf{B}^2 + \mathbf{x} (H-E) \cdot \mathbf{B} + \mathbf{B} \cdot \mathbf{x}
  (H-E)  + \mathbf{x} (H-E) \cdot \mathbf{x} (H-E),
\end{equation}
where an explicit expression for $\mathbf{B}^2$ is already known, as it is given by (\ref{eq:B^2}).\par
%
%
{}From
\begin{equation}
  \mathbf{x} (H-E) \cdot \mathbf{x} (H-E) = r^2 (H-E)^2 - {\rm i} \mathbf{x} \cdot \mathbf{p} (H-E),
\end{equation}
where
\begin{align}
  (H-E)^2 &= \frac{1}{4} \mathbf{p}^4 + \alpha \frac{\boldsymbol{\gamma} \cdot \mathbf{x}}{r^2}
       \left\{\mathbf{p}^2 + \frac{{\rm i}}{r^2} [\mathbf{x} \cdot \mathbf{p} - {\rm i}(d-2)] + \frac{1}{r^2}
       L_{ij} S_{ij} \right\} \nonumber \\
  &\quad {}+ \frac{\alpha^2}{r^2} - E \left(\mathbf{p}^2 + 2\alpha \frac{\boldsymbol{\gamma} \cdot \mathbf{x}}
       {r^2}\right) + E^2. \label{eq:C.3}
\end{align}
and
\begin{equation}
  - {\rm i} \mathbf{x} \cdot \mathbf{p} (H-E) = - \frac{{\rm i}}{2} (\mathbf{x} \cdot \mathbf{p}) \mathbf{p}^2
  - \alpha {\rm i} \frac{\boldsymbol{\gamma} \cdot \mathbf{x}}{r^2} (\mathbf{x} \cdot \mathbf{p} + {\rm i})
  + E {\rm i} \mathbf{x} \cdot \mathbf{p},
\end{equation}
we obtain`
\begin{align}
  &\mathbf{x} (H-E) \cdot \mathbf{x} (H-E) \nonumber \\
  & \quad = \frac{1}{4} r^2 \mathbf{p}^4 - \frac{{\rm i}}{2} (\mathbf{x}
       \cdot \mathbf{p}) \mathbf{p}^2 + \alpha (\boldsymbol{\gamma} \cdot \mathbf{x}) \left(\mathbf{p}^2
       + \frac{d-1}{r^2} + \frac{1}{r^2} L_{ij} S_{ij}\right) \nonumber \\
  & \qquad {}+ \alpha^2 + E \left(- r^2 \mathbf{p}^2 + {\rm i} \mathbf{x} \cdot \mathbf{p} - 2\alpha
       \boldsymbol{\gamma} \cdot \mathbf{x}\right) + E^2 r^2.
\end{align}
Note that in deriving (\ref{eq:C.3}), we used (\ref{eq:B.2}), as well as the relation
\begin{equation}
  - {\rm i} \boldsymbol{\gamma} \cdot \mathbf{p} = \frac{\boldsymbol{\gamma} \cdot \mathbf{x}}{r^2} 
  (- {\rm i} \mathbf{x} \cdot \mathbf{p} + L_{ij} S_{ij}),
\end{equation}
coming from (\ref{eq:relations}).\par
%
%
{}Furthermore,
\begin{align}
  &\mathbf{x} (H-E) \cdot \mathbf{B} + \mathbf{B} \cdot \mathbf{x} (H-E) \nonumber \\
  & \quad = (\mathbf{x} \cdot \mathbf{B} + \mathbf{B} \cdot \mathbf{x}) (H-E) + {\rm i} \mathbf{x} \cdot
         \mathbf{p} (H-E) \nonumber \\
  &  \quad = \left[2 \mathbf{x} \cdot \mathbf{B} + {\rm i} d \left(\mathbf{x} \cdot \mathbf{p} - {\rm i}
         \frac{d-1}{2}\right) + {\rm i} \mathbf{x} \cdot \mathbf{p}\right] (H-E) \nonumber \\
  &  \quad = [r^2 \mathbf{p}^2 - 2 (\mathbf{x} \cdot \mathbf{p})^2 + 2{\rm i} (d-1) \mathbf{x} \cdot 
         \mathbf{p} + L_{ij} S_{ij} + \tfrac{1}{2}d(d-1) + 2E r^2] \nonumber \\
  & \qquad {} \times (H-E),  \label{eq:C.7}
\end{align}
where use is successively made of (\ref{eq:LRL-proof}), (\ref{eq:commutator}), (\ref{eq:B-x}), and (\ref{eq:B}). The last relation is made of three terms. The first one is
\begin{align}
  &\tfrac{1}{2} [r^2 \mathbf{p}^4 - 2 (\mathbf{x} \cdot \mathbf{p})^2 \mathbf{p}^2 + 2{\rm i} (d-1)
      (\mathbf{x} \cdot \mathbf{p}) \mathbf{p}^2 + L_{ij} S_{ij} \mathbf{p}^2 + \tfrac{1}{2}d(d-1) 
      \mathbf{p}^2]\nonumber \\
  & {} + E r^2 \mathbf{p}^2.
\end{align}
\par
%
%
The second term can be written as
\begin{align}
  &\alpha [r^2 \mathbf{p}^2 - 2 (\mathbf{x} \cdot \mathbf{p})^2 + 2{\rm i} (d-1) (\mathbf{x} \cdot \mathbf{p})
      + L_{ij} S_{ij} + \tfrac{1}{2}d(d-1) + 2Er^2] \frac{\boldsymbol{\gamma} \cdot \mathbf{x}}{r^2}
      \nonumber \\
  & = \alpha \frac{\boldsymbol{\gamma}\cdot \mathbf{x}}{r^2} [r^2 \mathbf{p}^2 - 2 (\mathbf{x} \cdot 
       \mathbf{p})^2 + 2{\rm i} (d-1) (\mathbf{x} \cdot \mathbf{p}) + L_{ij} S_{ij} + \tfrac{1}{2}d(d-1) + 2Er^2]
       \nonumber \\
  & \quad {}+ \alpha r^2 \left[\mathbf{p^2}, \frac{\boldsymbol{\gamma} \cdot \mathbf{x}}{r^2}\right]
       - 2\alpha \left[(\mathbf{x} \cdot \mathbf{p})^2, \frac{\boldsymbol{\gamma} \cdot \mathbf{x}}{r^2}\right]
       + 2{\rm i} \alpha (d-1) \left[\mathbf{x} \cdot \mathbf{p}, \frac{\boldsymbol{\gamma} \cdot \mathbf{x}}{r^2}
       \right] \nonumber \\
  & \quad {}+ \alpha \left[L_{ij} S_{ij}, \frac{\boldsymbol{\gamma} \cdot \mathbf{x}}{r^2}\right] \nonumber \\
  & = \alpha \frac{\boldsymbol{\gamma} \cdot \mathbf{x}}{r^2} [r^2 \mathbf{p}^2 - 2 (\mathbf{x} \cdot 
       \mathbf{p})^2 + 2{\rm i} (d-2) \mathbf{x} \cdot \mathbf{p} + L_{ij} S_{ij} + \tfrac{1}{2}(d-1)(d-2)]
       \nonumber \\
   & \quad {}+ 2\alpha E (\boldsymbol{\gamma} \cdot \mathbf{x}),    
\end{align}
where use has been made of (\ref{eq:B.2}), (\ref{eq:B.3}), as well as
\begin{equation}
  \left[(\mathbf{x} \cdot \mathbf{p})^2, \frac{\boldsymbol{\gamma} \cdot \mathbf{x}}{r^2}\right] =
  \frac{\boldsymbol{\gamma} \cdot \mathbf{x}}{r^2} (2{\rm i}\mathbf{x} \cdot \mathbf{p} - 1)
\end{equation}
and
\begin{equation}
  \left[L_{ij} S_{ij}, \frac{\boldsymbol{\gamma} \cdot \mathbf{x}}{r^2}\right] = \frac{1}{r^2} \left[- 2{\rm i}
  (\boldsymbol{\gamma} \cdot \mathbf{x}) \left(\mathbf{x} \cdot \mathbf{p} - {\rm i} \frac{d-1}{2}\right)
  + 2{\rm i} r^2 (\boldsymbol{\gamma} \cdot \mathbf{p})\right],
\end{equation}
as a consequence of (\ref{eq:gamma}), (\ref{eq:S}), and (\ref{eq:B.6}).\par
%
%
Since the third term is simply
\begin{equation}
  -E [r^2 \mathbf{p}^2 - 2 (\mathbf{x} \cdot \mathbf{p})^2 + 2{\rm i}(d-1) \mathbf{x} \cdot \mathbf{p}
  + L_{ij} S_{ij} + \tfrac{1}{2}d(d-1)] - 2E^2 r^2,
\end{equation}
the result for (\ref{eq:C.7}) is
\begin{align}
  &\mathbf{x} (H-E) \cdot \mathbf{B} + \mathbf{B} \cdot \mathbf{x} (H-E) \nonumber \\
  &= \tfrac{1}{2} [r^2 \mathbf{p}^4 - 2 (\mathbf{x} \cdot \mathbf{p})^2 \mathbf{p}^2 + 2{\rm i} (d-1)
      (\mathbf{x} \cdot \mathbf{p}) \mathbf{p}^2 + L_{ij} S_{ij} \mathbf{p}^2 + \tfrac{1}{2}d(d-1) 
      \mathbf{p}^2]\nonumber \\
  & \quad {}+ \alpha \frac{\boldsymbol{\gamma} \cdot \mathbf{x}}{r^2} [r^2 \mathbf{p}^2 - 2 (\mathbf{x} \cdot 
       \mathbf{p})^2 + 2{\rm i} (d-2) \mathbf{x} \cdot \mathbf{p} + L_{ij} S_{ij} + \tfrac{1}{2}(d-1)(d-2)]
       \nonumber \\
  & \quad {}+ E [2\alpha \boldsymbol{\gamma} \cdot \mathbf{x} + 2 (\mathbf{x} \cdot \mathbf{p})^2
       - 2{\rm i}(d-1) \mathbf{x} \cdot \mathbf{p} - L_{ij} S_{ij} - \tfrac{1}{2}d(d-1)] - 2E^2 r^2.
\end{align}
\par
%
%
The final result for $\mathbf{\tilde{A}}^2$ reads
\begin{align}
  \mathbf{\tilde{A}}^2 &= r^2 \mathbf{p}^4 - (\mathbf{x} \cdot \mathbf{p})^2 \mathbf{p}^2 + {\rm i}(d-2)
       (\mathbf{x} \cdot \mathbf{p}) \mathbf{p}^2 + L_{ij} S_{ij} \mathbf{p}^2 + \tfrac{1}{4}d(d-1) \mathbf{p}^2
       \nonumber \\
  & \quad {}+ 2 \alpha \frac{\boldsymbol{\gamma} \cdot \mathbf{x}}{r^2} [r^2 \mathbf{p}^2 - (\mathbf{x}
       \cdot \mathbf{p})^2 + {\rm i}(d-2) \mathbf{x} \cdot \mathbf{p} + L_{ij} S_{ij} + \tfrac{1}{4}d(d-1)]
       \nonumber \\
  & \quad {}+ \alpha^2,
\end{align}
which, on comparing with (\ref{eq:term-1}), leads to equation (\ref{eq:relation-1}).\par
%
%

\end{document}